# Generalized Entropy Optimized by

# an Arbitrary Distribution

Sumiyoshi Abe

*Institute of Physics*, *University of Tsukuba*, *Ibaraki 305-8571*, *Japan*

We construct the generalized entropy optimized by a given arbitrary statistical distribution with a finite linear expectation value of a basic random quantity of interest. This offers, via the maximum entropy principle, a unified basis for a great variety of distributions observed in nature, which can hardly be described by the conventional methods. As a simple example, we explicitly derive the entropy associated with the stretched exponential distribution. To include the distributions with the divergent moments (e.g., the Lévy stable distributions), it is necessary to modify the definition of the expectation value.

Diversity of statistical distributions observed in nature is truly remarkable. In particular, a number of distributions, which are anomalous in view of ordinary statistical mechanics, can be recognized in a variety of complex systems. To understand the properties of such complex systems deeper, it is desirable to characterize these distributions within a unified framework of the statistical principles. The maximum entropy principle can be thought of as such one [1].

In a physical experiment, what to be measured is quite often the distribution and not directly the entropy itself. Therefore, it is necessary to be able to find the corresponding entropic measure optimized by the observed distribution under appropriate constraints.

In this paper, we present the entropy-generating algorithm and construct the generalized entropy, which can be optimized by a given arbitrary distribution with the finite *linear expectation value* of a random quantity of interest. As an example, we explicitly derive the entropy associated with the stretched exponential distribution.

Let us start our discussion by considering a continuous function $f$ of $t \in D \subseteq \mathbf{R}$, whose range is $[0, 1]$ and is assumed to be integrable over $D$

$$F = \int_D dt\, f(t) < \infty. \tag{1}$$

$f$ need not be a monotonic function, in general.

In terms of a probability distribution $\{p_i\}_{i=1, 2, \cdots, W}$ with $W$ the number of accessible states, we define the following quantity:

$$A[p;t] = \sum_{i=1}^{W} \left(p_i - f(t)\right)_+. \qquad (2)$$

In this equation, the symbol $(x)_+$ stands for

$$(x)_+ = \max\{0, x\}, \qquad (3)$$

which satisfies

$$(\lambda x + (1-\lambda) y)_+ \le \lambda (x)_+ + (1-\lambda)(y)_+ \qquad (\forall \lambda \in [0,1]), \qquad (4)$$

$$(x)_+ = x\,\theta(x), \qquad (5)$$

where $\theta(x)$ is the Heaviside unit step function defined by $\theta(x) = 0$ for $x < 0$ and $\theta(x) = 1$ for $x > 0$. Though we are considering a discrete distribution, generalizations of the subsequent results to the continuous case are straightforward. In the case when the distribution is continuous, the range of $f$ should be extended from $[0,1]$ to $[0,\infty)$, in general.

The quantity in Eq. (2) has some interesting properties. Among others, what we notice here are the following two:

$$0 < A[p;t] < 1, \qquad (6)$$

$$A[\lambda p + (1-\lambda) p'; t] \leq \lambda A[p; t] + (1-\lambda) A[p'; t] \qquad (\forall \lambda \in [0,1]). \qquad (7)$$

The latter directly follows from Eq. (4).

Next, we consider the integral

$$I[p] = \int_D dt \left(1 - A[p; t]\right), \qquad (8)$$

which is manifestly a positive functional due to Eq. (6). Using the expression in Eq. (5), we find this integral to be rewritten as follows:

$$I[p] = \sum_{i=1}^{W} \int_D dt \left(p_i - f(t)\right)\left[1 - \theta\left(p_i - f(t)\right)\right] + W F, \qquad (9)$$

where $F$ is given in Eq. (1).

Our entropy-generating algorithm consists of essentially identifying the generalized entropy, $S$, with the positive functional $I$:

$$S[p] = k I[p] + a, \qquad (10)$$

where $k$ and $a$ are constants, and in particular $k$ has to be positive. Since the entropy should vanish for the completely ordered state, $p_i = p_i^{(0)} = \delta_{in}$ (with $n$ a natural number between 1 and $W$), the constant $a$ satisfies the condition: $k I[p^{(0)}] + a = 0$.

Consequently, the generalized entropy is given by

$$S[p] = k\left(I[p] - I[p^{(0)}]\right). \tag{11}$$

Note that this is, in fact, a concave functional:

$$S[\lambda p + (1-\lambda) p'] \geq \lambda S[p] + (1-\lambda) S[p'], \tag{12}$$

as can be ascertained by using Eq. (7).

Let us first discuss a relatively simple case when $f$ is a monotonically decreasing function defined in $D = [0, \infty)$. In this case, $I$ in Eq. (9) can further be written as

$$I[p] = \sum_{i=1}^{W} \left[ p_i f^{-1}(p_i) - \int_0^{f^{-1}(p_i)} dt\, f(t) \right] + W F, \tag{13}$$

where $f^{-1}$ is the inverse function of $f$. Therefore, the generalized entropy reads

$$S[p] = k\left\{ \sum_{i=1}^{W} \left[ p_i f^{-1}(p_i) - \int_0^{f^{-1}(p_i)} dt\, f(t) \right] + c \right\}, \tag{14}$$

where $c$ is a constant given by

$$c = W \int_{0}^{f^{-1}(0)} dt\, f(t) + \int_{f^{-1}(0)}^{f^{-1}(1)} dt\, f(t) - f^{-1}(1). \tag{15}$$

Let us employ $S$ in eq. (14) as the entropy for the maximum entropy principle. Consider the functional

$$\Phi[p:\alpha,\beta] = S[p] - \alpha\left(\sum_{i=1}^{W} p_i - 1\right) - \beta\left(\sum_{i=1}^{W} p_i Q_i - <Q>\right), \tag{16}$$

where $Q_i$ is the $i$th value of the basic random variable, $Q$, and $\alpha$ and $\beta$ are the Lagrange multipliers associated with the constraints on the normalization condition and on the linear expectation value of $Q$ denoted by $<Q>$, respectively. (Here, clearly the ordinary expectation value is assumed to be well defined. However, there exist distributions with no finite moments. Celebrated examples are the Lévy stable distributions. To treat such distributions, it is necessary to modify the definition of the expectation value. See the later remarks.) Under the assumption of differentiability of $f^{-1}$, variation of $\Phi$ with respect to $\{p_i\}_{i=1,2,\cdots,W}$ leads to the following stationary distribution:

$$p_i = f(\alpha + \beta Q_i), \tag{17}$$

provided that the positive constant, $k$, has been eliminated by rescaling the Lagrange multipliers. $\alpha$ can be determined by the normalization condition:

$\sum_{i=1}^{W} f(\alpha + \beta Q_i) = 1$. Therefore, an arbitrary monotonically decreasing distribution with the finite linear expectation value of $Q$ could be derived from the maximum entropy principle.

The above discussion can immediately be applied to the stretched exponential distribution. In this case, $f$ is taken to be

$$f(t) = \exp(-t^\gamma) \qquad (t \in [0, \infty), \ \gamma \in (0, 1)). \tag{18}$$

Then, the associated generalized entropy is found to be given by

$$S[p] = \sum_{i=1}^{W} \Gamma(1 + 1/\gamma, -\ln p_i) - \frac{1}{\gamma} \Gamma(1/\gamma), \tag{19}$$

where $\Gamma(s, x)$ is the incomplete gamma function of the second kind defined by $\Gamma(s, x) = \int_{x}^{\infty} dt \, t^{s-1} e^{-t}$, and $\Gamma(s) = \Gamma(s, 0)$ is the gamma function. The generalized entropy in Eq. (19) is identical to the one recently discussed in Ref. [2], in which it seems to have come from the sky.

We mention that, taking the limit $\gamma \to 1 - 0$ in the above discussion, $f$ becomes the exponential function of $t$ and Eq. (19) converges to the ordinary Boltzmann-Gibbs-Shannon entropy, as it should do [3]. It can be shown [4] that if $f$ is taken to be the $q$-exponential function [5,6], then the corresponding generalized entropy is the Tsallis entropy [7]. Similarly, the so-called $\kappa$-deformed entropy [8,9] can also be derived if the $\kappa$-deformed exponential function is employed (though we do not present its explicit derivation here).

Now, let us succinctly discuss the case when $f$ in Eq. (2) is not monotonic. In such a case, the domain interval $D$ of $f$ should be divided into the subintervals, $\{D_a\}$, in which $f$ is monotonic. $D$ is now the disjoint union of $D_a$'s. Accordingly, Eqs. (8) and (9) are written as

$$I[p] = \sum_a \int_{D_a} dt\, (1 - A[p;t]), \tag{20}$$

and

$$I[p] = \sum_a \sum_{i=1}^{W} \int_{D_a} dt\, (p_i - f(t))\left[1 - \theta(p_i - f(t))\right] + WF, \tag{21}$$

respectively. Let $D_a^* = D_a^*\left(f^{-1}(p_i)\right)$ be the subinterval of $D_a$, in which $p_i < f(t)$. Then, Eq. (21) can further be rewritten as follows:

$$I[p] = \sum_a \sum_{i=1}^{W} \int_{D_a^*} dt\, (p_i - f(t)) + WF$$

$$= \sum_a \sum_{i=1}^{W} \left[p_i\, l\left(D_a^*\left(f^{-1}(p_i)\right)\right) - F_a^*\left(f^{-1}(p_i)\right)\right] + WF, \tag{22}$$

where $l(D_a^*)$ is the length of the interval $D_a^*$ and $F_a^* = \int_{D_a^*} dt\, f(t)$. Substitution of this quantity into Eq. (11) leads to the generalized entropy optimized by a non-monotonic distribution.

Finally, we mention that the Legendre transform structure highlighted by the relation

$$\frac{\partial S}{\partial <Q>} = \beta \qquad (23)$$

exists in the present generalized theory. This is due to the fact [10,11] that Eq. (23) holds for an arbitrary form of the entropy and an arbitrary definition of the expectation value.

In conclusion, we have constructed the generalized entropy, which is optimized by any given statistical distribution. This can be regarded as an extreme generalization of the maximum entropy principle. The present approach assumes that the distribution has a finite linear expectation value of a basic random quantity of interest. If the lowest moments are divergent, then the definition of the expectation value should be modified. For example, in Tsallis statistics [3-5], the optimal distribution is the so-called *q*-exponential distribution, which can asymptotically be a power-law distribution. In this case, the expectation value should be defined in terms of the escort distribution [12]. Regarding necessity of modifying the definition of the expectation value, the approaches from the different perspectives are needed [13].